\documentclass[smallextended]{svjour3}       
\smartqed  
\usepackage{graphicx}
\usepackage{amsmath,amsfonts}
\usepackage{amssymb}

\usepackage{enumerate}
\usepackage{natbib}

\usepackage{graphicx,psfrag,epsf,epstopdf}
\usepackage[svgnames]{pstricks}

\usepackage[bottom]{footmisc}
\usepackage{url}

\usepackage{color}
\usepackage{caption}
\usepackage{booktabs}
\usepackage{bbm}

\newtheorem{prop}{Proposition}

\def\bmath0{{\boldmath 0}}
\def\bm1{{\boldsymbol 1}}
\def\bmb{{\boldsymbol b}}

\def\bmx{{\boldsymbol x}}
\def\bmy{{\boldsymbol y}}

\def\bmp{{\boldsymbol p}}

\def\bmz{{\boldsymbol z}}

\def\bmX{{\boldsymbol X}}
\def\bmY{{\boldsymbol Y}}

\def\bmq{{\boldsymbol q}}

\def\argmin{{\it argmin}}

\def\Var{{\it Var}}
\def\Cov{{\it Cov}}

\def\conv{{\emph{conv}}}

\def\IR{{\mathbb{R}}}
\def\R{\mathbb{R}}

\journalname{Journal of Economic Inequality}

\begin{document}

\title{Representative endowments and uniform Gini orderings of multi-attribute welfare
}

\titlerunning{Endowments and orderings of multi-attribute welfare}        

\author{Karl Mosler
}

\institute{Karl Mosler \at Institute of Econometrics and Statistics,
              University of Cologne \\ \email {kmosler@uni-koeln.de}
}

\date{Received: date / Accepted: date}

\maketitle

\begin{abstract}
For the comparison of inequality and welfare in multiple attributes the use of generalized Gini indices is proposed.
Individual endowment vectors are summarized by using attribute weights and aggregated in a spectral social evaluation function. Such functions are based on classes of spectral functions, ordered by their aversion to inequality. Given a spectrum and a set $P$ of attribute weights, a multivariate Gini dominance ordering, being uniform in weights, is defined. If the endowment vectors are comonotonic, the dominance is determined by their marginal distributions; if not, the dependence structure of the endowment distribution has to be taken into account.
For this, a set-valued representative endowment is introduced that characterizes the welfare of a $d$-dimensioned distribution.
It consists of all points above the lower border of a convex compact in $\R^d$, while the
set ordering of representative endowments corresponds to uniform Gini dominance.
An application is given to the welfare of 28 European countries.
Properties of $P$-uniform Gini dominance are derived, including relations to other orderings of $d$-variate distributions such as convex and dependence orderings.
The multi-dimensioned representative endowment can be efficiently calculated from data. In a sampling context, it consistently estimates its population version.


\end{abstract}

\section{Introduction}
\label{sec:1}

Socio-economic status has many aspects. It is not only income on which the welfare of a person is based. Other dimensions of well-being such as wealth or education substantially influence the position of an individual and thus the degree of welfare in a society. Inequality and welfare in several, say $d$ dimensions, can be measured by indices and orderings that operate on distributions in $\IR^d$, where each individual of a population is represented by her or his vector of endowments.
An early source of such multivariate approaches is \cite{Kolm77}; for a recent survey see \cite{AndreoliZ20}.

The classical uni-dimensional  Gini index measures income inequality by the difference between mean income and a weighted mean of incomes, $\overline{x} -  \sum_{i=1}^n x_{(i)} (2n - 2i + 1)/{n^2}$, where $x_{(1)}, \dots, x_{(n)}$ are incomes ordered from below, and $\overline{x}$ is their mean.
It has been extended to the generalized Gini index (\cite{Mehran76}),
\begin{equation}\label{GenGinabs}
  G(x_1,\dots, x_n)= \overline{x} - \sum_{i=1}^n x_{(i)} w_i\,,
\end{equation}
incorporating general weights $w_1\ge w_2\ge \dots \ge w_n\ge 0$ that sum up to $1$. For an axiomatic characterization of (\ref{GenGinabs}), see \cite{Weymark81}. The term $\sum_{i=1}^n x_{(i)} w_i$ is mentioned as the \textit{representative income} of the distribution. It takes its maximum if all weights $w_i$ are equal, yielding $G(x_1,\dots, x_n)=0$. The representative income reflects two aspects of the distribution: the general level of incomes as well as the inequality among them.
Note that (\ref{GenGinabs}) is an \textit{absolute Gini index}.
By assuming $\overline x > 0$ and substituting relative endowments $x_i/\overline{x}$ for the $x_i$, a \textit{relative Gini index} is obtained, which focusses on inequality of income shares.

To apply these indices for welfare comparisons, however, parameters have to be inserted, which often are not available.
To manage this, dominance relations have been introduced in the literature that hold for certain sets of parameters.
Representative incomes have been constructed on the basis of generalized Gini indices and distributions have been ordered by those representative endowments \citep{Weymark81}. Uncertainty about the degree of inequality aversion has been coped with by unanimous preference orderings.
\cite{MuliereS89} investigate a sequence of progressively finer stochastic orderings, called \textit{inverse stochastic dominance}, and use it to order income distributions regarding their inequality; see also \cite{LandoBB20}.
Inverse stochastic dominance employs the {quantile functions} of two random variables in the same way as classical stochastic dominance employs their distribution functions.
\cite{Weymark81} relates $n$-th inverse stochastic dominance to the {ordering of S-Gini indices} having a large enough exponent.

My goal is to address these tasks in the case of several attributes: to provide a framework of multi-dimensioned representative endowments, and to construct unanimous orderings under uncertainty about the relative weights of the attributes and also about the degree of inequality aversion.

Consider $d$-variate vectors of endowments in a population, which form the rows of an $n\times d$ matrix $\bmX=(x_i^j)$.
More general, we treat multivariate probability distributions resp.\ random vectors $\bmX$ in $\R^d$. An endowment matrix then corresponds to an empirical distribution on its rows.

Several authors propose indices of multivariate inequality, aggregating the information to a single number.
\cite{Tsui95} constructs an index that aggregates the data in two steps, first over individuals and then over attributes.
Similarly, \cite{GajdosW05} derive a multivariate generalized Gini index from axioms including those of comonotonic independence and attribute separability.
In their dual theory of multivariate risk, \cite{GalichonH12} present a general risk index, which does not assume attribute separability and is based on multivariate quantiles; if interpreted as an inequality index it comes out as an extension of the Gajdos-Weymark index.
In contrast, the multivariate Gini index of \cite{KoshevoyM97} measures the mean Euclidean distance of the $d$-variate endowments from each other.

In the sequel we start with social evaluation functions for one attribute, say income, which are based on integrals involving a distorted distribution.
The representative endowment is a spectral measure of the distribution, the spectrum being the derivative of a given distortion (i.e.\ weight-generating) function; the generalized Gini measure comes as the difference of the representative income to the mean income. Concavity of the distortion function corresponds to inequality aversion. The representative endowment decreases if the distortion function is changed to a more concave ( = less convex) one.
Parametric classes of distortion functions are considered that show increasing inequality aversion.
Special cases are discussed where the representative endowment is the highest income of the lower $\alpha \cdot 100$ percentage of the population or the mean income of this subpopulation. Given a distortion function, any two endowment distributions are compared by their representative incomes. This provides a complete ordering of distributions that reflects their inequality as well as their general level.
In case of uncertainty about the
distortion weights to be put on incomes, a set of admissible distortion functions may be considered. Uniform comparison of representative endowments regarding a set of distortion functions yields a partial order of distributions. For certain such sets it comes out that the respective ordering is equivalent to known stochastic orders, \emph{viz.} first and second degree stochastic dominance relations as well as higher degree dual stochastic dominance orders.

With several attributes we introduce a vector $\bmp\in P_+$ of attribute weights by which the $d$ attribute levels are aggregated,  $P_+$ being the standard simplex in $\IR^d_+$. These weights reflect the relative importance of the attributes. Each attribute is valued by a (generally unknown) weight or `price', and the welfare of the `priced endowment' is assessed.
Given a distortion function, this allows for any $\bmp$ to build a representative `priced endowment' and to produce a generalized Gini index. Here contrary to \cite{Tsui95} and successors the aggregation is first done over the attributes and then over the population.
As the weights are generally unknown, a partial order of distributions is defined, \emph{$P$-uniform Gini dominance}, saying that two representative `priced endowments' are ordered in the same way for some set $P$ of non-negative price vectors $\bmp$, which reflects the available information about $\bmp$.
This in particular incorporates that certain attributes may have no common unit of measurement.

It is shown that, if each endowment vector $\bmX$ and $\bmY$ is comonotonic, $P_+$-uniform Gini dominance is equivalent to dominance of the marginal distributions. If not, the dependence structure of the multivariate distribution of endowments has to be taken into account.
To cope with this, we construct a visual tool by which two distributions can be compared.
The welfare of a multi-attribute distribution, given a distortion function, is geometrically characterized by a convex upper set in $d$-space, the \textit{convex representative endowment (CRE)}. It again is described by its set of minimal ( = Pareto) points, which is called the \textit{Pareto representative endowment (PRE)}. For each price vector $\bmp$, the PRE contains a point $\bmx_\bmp$ that lies on the ray in direction $\bmp$, and $\bmx_\bmp$ serves as the equally distributed representative vector of endowments under the given price vector.
Every point in CRE is not worse than $\bmx_\bmp$, and the CRE comes out as the union of upper orthants of all $\bmx_\bmp$.
Uniform Gini dominance regarding all $\bmp \ge 0$ comes out to be equivalent to set inclusion of the respective CREs.

The definitions, so far, assume a fixed underlying distortion function. Next, as in the univariate case, classes of distortion functions are discussed which are ordered by their inequality aversion. A less convex, that is more inequality averse distortion function
enlarges the CRE.

$P$-uniform Gini dominance relations will be shown to be increasing orders as well as orderings of variability. For any concave ( = inequality averse) distortion function and any information $P$ on prices, they are implied by usual multivariate stochastic order as well as reverse
 convex stochastic order. Moreover, weak price majorization proves to be stronger than $P$-uniform Gini dominance. Further, $P$-uniform Gini dominance is shown to be related to increasing correlation; it is implied by the submodular order. Finally, $P$-uniform Gini dominance relations are introduced which are uniform in a given set of distortion functions.

When it comes to data, the PRE can be numerically calculated by existing software. By this, two empirical distributions are numerically checked for $P$-uniform Gini dominance. If the data is sampled from an underlying distribution, the empirical PRE is a consistent estimator of its underlying population version.

Overview: Section 2 treats generalized Gini orderings and representative endowments in the univariate case. Uniform Gini dominance with multiple attributes is introduced in Section 3 and the comonotonic case is treated. Section 4 presents the convex representative endowment and its Pareto set,
while Section 5 is about increasing inequality aversion and an application to well-being data of 28 European countries.
Section 6 shortly discusses a similar scale invariant measurement of multi-attribute inequality.
In Section 7 uniform Gini dominance is related to known multivariate orderings of variability and dependence.
Section 8 concludes with remarks on statistical and computational issues.
Most proofs are collected in an Appendix.

\section{Gini orderings and representative endowments in one dimension}\label{sec:2}
We start with a single dimension of welfare, say income, and discuss social evaluation functions which are based on an integral that involves a distorted distribution.

Let $X$ be a variable of socio-economic interest, which is modeled as a real-valued random variable on a proper probability space, having distribution function $F_X$.
In particular, $X$ may have an {empirical distribution},
giving equal probability $\frac 1n$ to points $x_1,\dots, x_n\in \mathbb{R}$.

Let $W_0$ be the set of all functions $v:[0,1]\to [0,1]$ that are
 \textit{non-decreasing},  \textit{right-continuous}, and satisfy
$v(0)=0\,, \ v(1)=1$. In other words, $W_0$ contains the probability distribution functions living on $[0,1]$. Given $v\in W_0$, we define its \textit{dual function} $\tilde v$ as $\tilde v(t) = 1- v(1-t)$.
For $v\in W_0$ consider the integral
\begin{equation}\label{eq:spectralevaluation}
  S_v(X) =  \int_0^1  Q_X(t) dv(t)\,,
\end{equation}
where $Q_X(t)= \min \{ x : F_X(t)\ge t\}\,, \; t\in ]0,1],$
denotes the usual left continuous inverse of $F_X$.
The integral (\ref{eq:spectralevaluation}) becomes
\begin{eqnarray}\label{eq:spectralevaluation2}
S_v(X)&=& \int_{-\infty}^\infty x \, d(v\circ F_X(x)) \nonumber\\
&=& \int_0^\infty \tilde v(1-F_X(x))dx  - \int^0_{-\infty} v(F_X(x))dx \,. \nonumber
\end{eqnarray}
(Here and in the sequel we tacitly assume that these integrals exist and are finite, if necessary, under proper regularity conditions.)
Obviously, for $v(t) = \tilde v(t) = t$ the mean, $S_v(X)= E[X]$, is obtained. When $v$ has  a derivative $v'$,
then
\begin{equation}\label{eq:spectralevaluation1}
  S_v(X) =  \int_0^1  Q_X(t) v'(t) dt\,.
\end{equation}
$S_v(X)$ is a spectral measure of endowment having \textit{weight-generating function} $v$ and \textit{spectrum} $v'$. The function $v$ is called a \textit{distortion function} as it distorts the probabilities of results arising from $X$ in a generally nonlinear way. With other words, $S_v(X)$ is a weighted mean of endowments, which serves as a {\textit{representative endowment}}, being the aggregate evaluation of incomes weighted by their relevance to welfare.
Observe that the representative endowment of a distribution that gives everybody the same constant income equals the constant, hence $S_v(S_v(X))= S_v(X)$. Thus,
the representative endowment, if equally distributed in the population, yields the same welfare as the distribution of $X$ and is therefore mentioned as the \emph{equally distributed equivalent income}.
Usually, to measure the degree of welfare, lower incomes are given more weight than higher ones, that is, the distortion function $v$ is assumed to be concave. Then the spectral measure $S_v(X)$ is called \emph{inequality averse}, and the representative income undermatches the mean income, $S_v(X)\le E[X]$.

For an empirical distribution on $x_1,\dots,x_n$, one obtains a weighted mean of the ordered data,
\begin{equation*}\label{-S-empirical}
    S_v(X)= \sum_{i=1}^n x_{(i)} w_i\,, \quad \text{with} \;\; w_i=v\left(\frac in\right)-v\left(\frac{i-1}n\right)\,.
\end{equation*}
The \textit{generalized Gini index} $G(X)$ compares $S_v(X)$ with the non-weighted mean $E[X]$,
\begin{equation}\label{genGini}
  G(X)=E[X]-S_v(X)\,,
\end{equation}
which for empirically distributed $X$ is Mehran's index (\ref{GenGinabs}).
Choosing $v_\beta(t) = 1-(1-t)^\beta$, $\beta\ge 1$, produces the famous \textit{S-Gini index} introduced by \cite{DonaldsonW80}, \cite{DonaldsonW83} and \cite{Yitzhaki83},
\begin{equation}\label{S-GiniIndex}
G_{\beta}(X)= E[X]-S_\beta(X) = \int_0^1  Q_X(t) [1- \beta(1-t)^{\beta-1}] dt\,.
\end{equation}
Here, the parameter $\beta$ indicates the {\textit{aversion towards inequality}}:
With increasing $\beta$, lower values of $X$ get more weight. While (\ref{S-GiniIndex}) is an absolute S-Gini index, the relative index is given by $G_{S_\beta}(X)/E[X]$, provided $E[X]>0$.

With an empirical distribution on $x_1,\dots, x_n$ it holds
\begin{equation*}
S_\beta(x_1,\dots, x_n) = \sum_{i=1}^n x_{(i)}\int_{\frac{i-1}n}^{\frac in}\beta(1-t)^{\beta-1}dt
 = \sum_{i=1}^n x_{(i)} \left(\frac{n-i+1}n\right)^\beta- \left(\frac{n-i}n\right)^\beta \,.
\end{equation*}
Especially $\beta=1$ yields $S_1(x_1,\dots, x_n)=\overline x$\,.
For $\beta=2$ obtain
\begin{eqnarray*}
S_2(x_1,\dots, x_n) &=& \frac 1{n^2} \sum_{i=1}^n x_{(i)} (2n-2i+1)\,, \\
G_{2}(x_1,\dots, x_n) &=& \overline x - S_2(x_1,\dots, x_n) = \frac 1{n^2} \sum_{i=1}^n x_{(i)} (2i-n-1)\,,
\end{eqnarray*}
the classical absolute \textit{Gini index} \citep{Gini12}.

In the sequel further parameterized families of weight-generating (= distortion) functions are considered,
whose parameter $\alpha\in ]0,1]$ indicates the degree $\frac 1\alpha$ of inequality aversion. With
\begin{equation}\label{salpha}
s_{\alpha}(t)=\left\{\begin{array}{cl}
0\quad &\text{if $t<\alpha$,}\\
1 \quad &\text{if $t\ge \alpha$,}
\end{array}\right.
\end{equation}
obtain $S_{s_\alpha}(X)= Q_X(\alpha)$, the {\textit{highest income of the lower $\alpha$-part}} of the population.

Another interesting family is given by
\begin{equation}\label{ralpha}
r_{\alpha}(t)=\left\{\begin{array}{cl}
\frac t\alpha \quad &\text{if $t<\alpha$,}\\
1 \quad &\text{if $t\ge \alpha$ .}
\end{array}\right.
\end{equation}

This yields $S_{r_\alpha}(X)= \frac 1\alpha \int_{]-\infty,Q_X(\alpha)]} x \,  dF_X(x)$,
{\textit{the mean income of the lower $\alpha$-part}} of the population. Note that the $r_\alpha$ are concave, while the $s_\alpha$ are not.

\cite{MaccheroniMZ05} have introduced {partial orderings} of income distributions based on their representative incomes over a {range of weight-generating functions} as follows.
Each element of $W_0$ defines a rank-dependent weighting scheme by which incomes enter the social evaluation. Let $V$ be a subset of $W_0$ and define:
A random variable $Y$ dominates another random variable $X$ in \textit{$V$-dual stochastic dominance},
{$X\preceq_V Y$}, shortly \textit{$V$-dominance}, if
\begin{equation}\label{V-dom}
\int_0^1  Q_X(t) dv(t) \le \int_0^1  Q_Y(t) dv(t) \quad \text{for all} \;\; v\in V\,.
\end{equation}
If $V=\{v\}$ is a singleton, we write $X\preceq_v Y$ in place of $X\preceq_V Y$.
Note that $\preceq_V$ is a preorder (reflexive and transitive), but not necessarily antisymmetric. E.g., when $V = \{r_\alpha : \alpha < \alpha^*\}$ with $r_\alpha$ as in (\ref{ralpha}) and some $\alpha^*<1$, only the quantile restricted left tails of the two distributions are considered, which means that the welfare measurement focuses on the lower income population.

With special choices for the class $V$ of weight-generating functions, well known stochastic dominance (SD) relations arise:
\begin{itemize}
  \item  Let $V_1=W_0$.
$X\preceq_{V_1} Y$ is equivalent to usual \textit{first degree SD}, that is
\[F_X(z)\ge F_Y(z)\quad \text{for all}\;\; z\in \IR\,.\]
  \item Consider $V_{conc}=\{v\in W_0 : v\;\; \text{concave}, v' \; \text{bounded}\}$.
  Then
$X\preceq_{V_{conc}} Y$ is equivalent to usual \textit{second degree concave SD}, that is
\[\int_{-\infty}^z F_X(x)dx \ge \int_{-\infty}^z F_Y(y)dy \quad \text{for all}\;\; z\in \IR\,.\]
In this case $Y$ is less dispersed than $X$.
\item With $V_{conv}=\{v\in W_0 : v \; \text{convex}, v' \; \text{bounded}\}$,
$X\preceq_{V_{conv}} Y$ is equivalent to usual \textit{second degree convex SD}, that is
\[\int^{\infty}_z F_X(x)dx \ge \int^{\infty}_z F_Y(y)dy \quad \text{for all}\;\; z\,.\]
Here, $Y$ is more dispersed than $X$. Note that the two second degree dominance relations are linked by
\[ X\preceq_{V_{conv}} Y  \quad \Leftrightarrow \quad  -Y\preceq_{V_{conc}} -X\,.
\]
\end{itemize}

A further example of {$V$-dominance} is the \textit{Donaldson-Weymark dominance}:
Let $DW(A)=\{v  : v(t)= 1-(1-t)^{\beta}, \beta\in A\}$ for some bounded set $A\subseteq [1,\infty[$. Then $X\preceq_{DW(A)}Y$ says that, measured by the S-Gini index, the welfare  of $X$ is larger than that of $Y$ for every aversion degree $\beta\in A$. Since $V_{DW(A)}\subseteq V_{conc}$, Donaldson-Weymark dominance is weaker than second degree concave stochastic dominance,
\[ X\preceq_{V_{conc}} Y \quad \Rightarrow \quad X\preceq_{DW(A)} Y\,.
\]



Instead of $S_v(X)$ one may likewise regard the upper interval
$ [S_v(X),\infty[ $
as a {\textit{set-valued representation}} of income and compare those representations by set inclusion. With (\ref{salpha}),
$ \bigl[Q_X(\alpha), \infty \bigr[ $
is the {\textit{income range of the upper $(1-\alpha)$-part}} of the population, while with (\ref{ralpha})
\[ \left[\frac 1\alpha \int_{]-\infty,Q_X(\alpha)]} x\, dF_X(x), \infty \right[
\]
is the {\textit{range of incomes larger than the mean income of the lower $\alpha$-part}}. With multiple attributes we will employ a similar set-inclusion approach.

\section{Uniform Gini dominance in several dimensions}\label{sec:3}
So far, a single attribute of welfare, say income, has been examined. Now consider $d$ attributes, each being desirable as a `good'. Let $v$ be a weight-generating function from $V_{conc}$.  $\bmX$ denotes a random vector in $\IR^d$ depicting the distribution of $d$-variate endowments in a population. This includes the empirical case, where $\bmX$ has a distribution giving equal mass $1/n$ to $n$ points in $d$-space.
For $\bmp\in P_+= \{\bmp \in \mathbb{R}^d : \bmp\ge 0, \sum_{i=1}^n p_i =1 \}$  consider
\begin{equation}\label{multrepr}
S_v(\bmp'\bmX)=\int_0^1  Q_{\bmp'\bmX}(t) d v(t)\,,
\end{equation}
where $\bmp'\bmX$ denotes the inner product.
$\bmp$  may be seen as a vector of relative `prices', that is, importance weights of attributes.
For a given vector $\bmp$, (\ref{multrepr}) is an amount of \textit{representative `priced endowment'} or \emph{aggregate value}.
It measures the welfare in the population; the larger this amount, the larger is the welfare. As $v$ is concave, the measure $S_v(\bmp'\bmX)$ is inequality averse and always smaller or equal to mean priced endowment $E[\bmp'\bmX]$.
Thus, as in the single-dimensional case, the difference
\begin{equation}\label{multindex}
 E[\bmp'\bmX]-S_v(\bmp'\bmX)
\end{equation}
serves as an absolute measure of inequality, and $S_v(\bmp'\bmX)$ is an amount of individual aggregate value that, if equally distributed, would yield the same level of welfare as the given distribution. E.g., if $S_v(\bmp'\bmX)/E[\bmp'\bmX]=0.9$, only 90 percent of total aggregate value would be needed for the same welfare level under equal distribution. Or, if $S_v(\bmp'\bmY)/S_v(\bmp'\bmX)=1.05$, the distribution of $\bmY$ yields a five percent welfare gain over the distribution of $\bmX$.
Here, welfare is defined in terms of individual aggregate value only. This does include a possible substitution of attribute levels: each individual can change her/his relative levels under the condition of a given aggregate value.

Observe that in (\ref{multrepr}) the individual data of wellbeing is
\begin{itemize}
  \item first aggregated {over the attributes} by an additive score,
  \item and then evaluated {over the population}.
\end{itemize}

Also \cite{Tsui95} constructs multivariate inequality indices that aggregate the data in two steps. But different from the present approach he aggregates
first over the attributes on an individual level and then over the population. The same is done in the index of \cite{GajdosW05}.

Two distributions, those of  $\bmX$ and $\bmY$, may be compared by the index (\ref{multrepr}) if some vector $\bmp$ of attribute weights is given. But in an applied setting, this will be rarely the case. Therefore, we introduce a uniform ordering that allows for uncertainty about $\bmp$. Let $\emptyset \not= P \subseteq P_+$ reflect some \emph{partial information} on the price vector $\bmp$.

\begin{definition}
For $v\in V_{conc}$ and $\emptyset \not= P\subseteq P_+$ define  ${\bmX\preceq_{vP} \bmY}$ if
\begin{equation}\label{uniformGini}
\int_0^1Q_{\bmp^\prime \bmX}(t) d v(t) \le   \int_0^1Q_{\bmp^\prime \bmY}(t) d v(t)\quad \text{for all} \ \bmp\in P\,,
\end{equation}
in words, $\bmY$ has \emph{higher  $v$-welfare} than $\bmX$ in \textit{$P$-uniform Gini dominance}.
If $P=P_+$, the relation is simply mentioned as \textit{uniform Gini dominance}.
\end{definition}


Often, partial information $P$ is described by a finite number of linear inequalities on $\bmp$, such as $p_1 + p_4 \le 0.8$, $p_1 + p_2 \le p_3$ and similar. Then $P$ is mentioned as a \emph{linear partial information on prices}. It implies that
its \emph{set of extremal points}, $Ext(P)$, is finite, $P=\conv(Ext(P))$ being the convex hull.
The following Proposition \ref{ExtremalP} says that, under linear partial information on prices and a certain requirement of comonotonicity, $P$-uniform Gini dominance can be checked by checking (\ref{uniformGini}) for a finite number of price vectors only, \emph{viz.} $\bmp\in Ext(P)$.

A set ${\cal U}$ of real random variables is called \emph{comonotonic} if there exists a random variable $V$ and for each $U\in{\cal U}$  a non-decreasing function $g_U$ such that
$U=_d g_U(Z)$ holds in distribution. Particularly, if ${\cal U}$ consists of the marginals of a given random vector $\bmX$, ${\cal U}= \{X_1, \dots, X_d\}$, the random vector is comonotonic in the usual sense.

\begin{prop}\label{ExtremalP} Assume that $P$ is a linear partial information on prices and each of the sets
$\{\bmp'\bmX : \bmp\in Ext(P)\}$ and $\{\bmp'\bmY : \bmp\in Ext(P)\}$ is comonotonic. Then
\[\bmX \preceq_{v P}\bmY \quad \Leftrightarrow \quad \bmX \preceq_{v Ext(P)} \bmY\,.\]
\end{prop}
For proof, see the Appendix A.
In case of no information on $\bmp$, $P=P_+$, we have $Ext(P) = \{(1,0,\dots,0)', \dots (0,\dots,0,1)'\}$. If $\bmX$ as well as $\bmY$ are comonotonic random vectors, the Proposition \ref{ExtremalP} implies that uniform Gini dominance is fully determined by dominance of their marginals:
\begin{corollary}\label{marginaldominance}
Let $\bmX$ as well as $\bmY$ be comonotonic random vectors. Then
\[\bmX \preceq_{v P_+}\bmY \quad \Leftrightarrow \quad X_i\preceq_{v}Y_i \;\; \text{for} \; i=1,2,\dots d\,.
\]
\end{corollary}

Of course, in general this is not the case. To establish P-uniform dominance, not only the marginals but also the dependence structures of the random vectors $\bmX$ and $\bmY$ have to be taken into account.
To cope with this, in what follows the set-valued representative endowment of a given random vector is constructed, which is a subset of $\IR^d$ reflecting the given inequality posture as well as the (total or partial) uncertainty about attribute weights.

\section{Convex compacts and representative endowments}\label{sec:Convex}
As a visual device to depict and compare the multivariate welfare of $\bmX$, {convex sets} in $d$-space are introduced
whose upper extension will serve as a convex representative endowment. Those sets are characterized by their support functions.

The \textit{support function} $h_K$ of a non-empty closed convex set $K\subseteq\IR^d$ is defined as
\[
{h_K:\IR^d\to\IR\cup \{\infty\}\,,\quad h_K(\bmp)=\max\{\bmp^\prime \bmx\,|\,\bmx\in K\}}\,.
\]
The support function gives, for each direction $\bmp\in\IR^d$, the {distance between the origin and
the tangent hyperplane} in outer direction $\bmp$; its values are finite if and only if $K$ is compact. A support function is convex and positive homogeneous of degree 1. On the other hand, every such function characterizes a unique closed convex set. Moreover, the inclusion of two closed convex sets, $K$ and $L$, is simply described by their support functions: $K\subseteq L$ if and only if $h_K(\bmp)\le h_L(\bmp)$ for all $\bmp \in \mathbb{R}^d$. For these and more properties, see e.g. \cite{Rockafellar70}.

Consider a function ${w\in V_{conv}}$, that is, $w$ convex, increasing and continuous with $w(0)=0$, $w(1)=1$, $w'$ being bounded. Assume that $E[||\bmX||]$ is finite. Then the function
\begin{equation}\label{supportconvex}
h(\bmp)=\int_0^1Q_{\bmp^\prime \bmX}(t)\,dw(t)\,,\quad \bmp\in \mathbb{R}^d\,,
\end{equation}
is obviously positive homogenous. It is also convex and takes always finite values \citep[Proposition 1]{DyckerhoffM10a}. Therefore, (\ref{supportconvex}) is the {support function} of a compact convex set, say $C(\bmX,w)$, in $\IR^d$.

Note that, given $\bmp$, the support function (\ref{supportconvex}) resembles the  representation  (\ref{multrepr}) of univariate welfare
but is based on a convex distortion function $w$ instead of a concave one.
If a distortion function $v$ is convex, its dual $\tilde v$ is concave, and {\it viceversa}. For any $\bmp\in \mathbb{R}^d$ it holds
\begin{equation}\label{dualnegative}
\int_{0}^{1} Q_{\bmp^\prime \bmX}(t)\,dv(t)= - \int_{0}^{1} Q_{- \bmp^\prime \bmX}(t)\,d\tilde v(t)\,.
\end{equation}
From Lemma \ref{lemmadualsupport} in Appendix A we conclude
\begin{equation}\label{dualsupport}
S_v(\bmp'\bmX)=  \int_0^1Q_{\bmp^\prime \bmX}(t)\,dv(t) = - h_{C(\bmX,\tilde v)}(-\bmp)\,, \quad \bmp\in \mathbb{R}^d\,.
\end{equation}
Thus, for $v\in V_{conc}$ and any $\bmp\in \mathbb{R}^d$, $S_v(\bmp'\bmX)$ equals the distance of a tangent hyperplane to
$C(\bmX,\tilde v)$ from the origin, namely the tangent hyperplane in outer direction $-\bmp$. $C(\bmX,\tilde v)$ is partially ordered by the usual componentwise ordering $\le$ of $\mathbb{R}^d$. Its set of minimal points, that is its Pareto minimum  $PRE(C(\bmX,\tilde v))$,
will serve us as a multivariate version of representative endowment.
As the attributes are `goods' and their prices are  non-negative,
we expand $C(\bmX,\tilde v)$ to its upper set,
\[C^+(\bmX,\tilde v)=C(\bmX,\tilde v) \oplus \IR^d_+\,.\]
Here, $\oplus$ means \textit{Minkowski addition} of sets,
$A\oplus B=\left\{a+b\, : \,a\in A, b\in B\right\}\,.$
Obviously, $C^+(\bmX,\tilde v)$ has Pareto minimum equal to that of $C(\bmX,\tilde v)$ and is uniquely characterized by its
support function
\begin{equation}\label{support+}
  h_{C^+(\bmX,\tilde v)}(-\bmp)=\begin{cases}
          h_{C(\bmX,\tilde v)}(-\bmp), & \mbox{if } \; \bmp\in \mathbb{R}^d_+\,, \\
          \infty, & \mbox{otherwise}\,.
        \end{cases}
\end{equation}
We define:
\begin{definition} \ $C^+(\bmX,\tilde v)=C(\bmX,\tilde v) \oplus \IR^d_+$
is the \emph{convex representative endowment (CRE)} of distribution $\bmX$ under the weight-generating function $v$.
Its Pareto minimum set is mentioned as the \emph{Pareto representative endowment (PRE)}.
\end{definition}
For $\bmp\ge 0$, consider
\[\bmx_\bmp \in \argmin\{\bmp'\bmx : \bmx \in C^+(\bmX,\tilde v)\}\,.\]
Observe that $C^+(\bmX,\tilde v)$ is the union of all upper orthants originating from $\bmx_\bmp$, $\bmp\ge 0$.
Every $\bmz\in C^+(\bmX,\tilde v)$ is not less desirable than $\bmx_\bmp$ given this price vector $\bmp$. Moreover it holds
$S_v(\bmp'\bmx_\bmp) = S_v(\bmp'\bmX)$, that is, $\bmx_\bmp$ is the equally distributed representative vector of endowments under price vector $\bmp$.

Hence, for two distributions of $\bmX$ and $\bmY$, it holds  $C^+(\bmY, \tilde v) \subseteq C^+(\bmX, \tilde v)$ if and only if $h_{C^+(\bmY, \tilde v)}(\bmp) \le h_{C^+(\bmX, \tilde v)}(\bmp)$ for all $\bmp\in \mathbb{R}^d_+$, equivalently, by (\ref{dualsupport}) and homogeneity, if \\
$\int_0^1 Q_{\bmp'\bmX}(t)\,d v(t) \le \int_0^1 Q_{\bmp'\bmY}(t)\,d v(t)$ whenever $\bmp\in P_+$\,. This again is tantamount saying that
$\bmX\preceq_{v P_+}\bmY$.  We have obtained the following geometrization of uniform Gini dominance:
\begin{theorem}\label{upperC}
Let $v\in V_{conc}$, $P=P_+$. Then
\[
\bmX\preceq_{v P_+} \bmY \quad \Longleftrightarrow \quad C^+(\bmY, \tilde v) \subseteq C^+(\bmX, \tilde v)\,.
\]
\end{theorem}
The inclusion of CREs is equivalent to uniform Gini dominance. Particularly in dimensions two and three this allows for a visual assessment of the ordering.
When the data is transformed by a positive affine-linear transformation, the convex representative endowment alters in the same way:
\begin{prop}\label{affinetransform}
Consider an $m\times d$ matrix $A$, $A\ge 0$, and $\bmb\in \IR^m$. Then
\begin{equation}\label{affineequivariant}
C^+(A\bmX+\bmb,\tilde v)=AC^+(\bmX,\tilde v)+\bmb\,.
\end{equation}
\end{prop}
For proof, see Appendix A.
As Proposition \ref{affinetransform} holds for a diagonal matrix having positive entries in the diagonal, we get:
\begin{corollary}\label{equivariant}
The CRE as well as the PRE of a $d$-variate distribution are translation and multivariate scale equivariant.
\end{corollary}
Further, with a projection matrix $A_J$, yielding $\bmX_J$ and $\bmY_J$, conclude:
\begin{corollary}\label{projection}
The CRE of a $d$-variate distribution restricted to attributes $j\in J\subseteq \{1,2,\dots, d\}$ equals the projection of the $d$-variate convex representative endowment to the subspace having indices  $j\in J$. The same holds for the PREs.
\end{corollary}
From Corollary \ref{projection} it is also clear that uniform Gini dominance persists if some of the attributes are dropped:
\begin{corollary}\label{ProjOrder}
For any $v\in V_{conc}$
\[
\bmX\preceq_{vP_+} \bmY \quad \Longrightarrow \quad \bmX_J\preceq_{vP_+} \bmY_J\\,,
\]
$\bmX_J$ and $\bmY_J$ being the subvectors regarding any subset $J\subseteq \{1,2,\dots, d\}$ of indices.
\end{corollary}

\section{Increasing inequality aversion}\label{Parameterizing}

For a single attribute, the S-Gini index $G_\beta$ and the related representative endowment provide a family of measures which is parameterized by the degree $\beta$ of aversion to inequality. Similarly, with multi-dimensioned endowments we introduce families of weight-generating functions depending on a parameter of inequality aversion.

Let $R=\{u_\alpha\}_{\alpha\in [0,1]}$ be a \emph{convex-ordered family of functions} from $V_{conc}$, that is, $u_\alpha$ becomes less convex with increasing $\alpha$ \citep{ChanPS90}.
Then, if $\alpha > \beta$, the graph of $u_\alpha$ lies above that of $u_\beta$. In other words, regarded as probability distribution functions, $u_\beta$ dominates $u_\alpha$ in first degree stochastic dominance.
Therefore with any $\bmp\in \mathbb{R}^d$ it follows that
\[ \int_0^1  Q_{\bmp'\bmX}(t) du_\alpha(t)  \le   \int_0^1  Q_{\bmp'\bmX}(t) du_\beta(t)\,,\]
since the integrand increases with $t$. With (\ref{support+}) and (\ref{dualsupport})
obtain:
\begin{prop}\label{parameteralpha}
Consider a convex-ordered family $R=\{u_\alpha\}_{0<\alpha\le 1} \subseteq V_{conc}$. Then
\[\alpha > \beta  \quad \Rightarrow \quad C^+(\bmX,\tilde u_\beta)\subseteq C^+(\bmX,\tilde u_\alpha)\,. \]
\end{prop}
For $\alpha > \beta$ the PRE of $\bmX$ evaluated with $\alpha$ lies below that evaluated with $\beta$.
In other words, the welfare measured by PRE decreases when the parameter of inequality aversion $\alpha$ increases.

As an example of a convex-ordered family, consider the \emph{Donaldson-Weymark class} $R_{DW}=\{w_\alpha\}_{0< \alpha\le 1}$,
\[
w_{\alpha}(t)=1- (1-t)^{1/\alpha}\,, \quad 0\le t \le 1\,,
\]
as used in the univariate S-Gini index (\ref{S-GiniIndex}) with $\beta=\frac 1\alpha$.
It yields $\tilde w_\alpha(t)=t^{1/\alpha}$ and the so called \textit{continuous ECH$^*$  regions} \citep{Cascos07}.
Note that \cite{GalichonH12} employ this weight generating function with $\alpha=1/2$.

Another example is given by the family $R_{zon}=\{r_\alpha\}_{0< \alpha\le 1}$ of concave distortion functions (\ref{ralpha})
\[
 r_{\alpha}(t)= \frac t \alpha \wedge 1  = \left\{\begin{array}{cl}
\frac t \alpha \,,&\text{if $0\le t\le\alpha$,}\\
      1 \,,&\text{if $\alpha<t\le 1$\,,}
\end{array}\right.
\]
hence $\tilde r_\alpha(t)=0\vee (t-1+\alpha)/\alpha$.
The resulting sets $C(\bmX,0\vee (t-1+\alpha)/\alpha)$ are mentioned as \emph{zonoid regions} (\cite{KoshevoyM97b}).
Observe that the family $R_{zon}$ consists of the `most concave' distortion functions, that is the minimal elements of the convex order among all distortion functions which are either convex or concave.

We illustrate the $R_{DW}$-based PREs by an example of welfare between countries, \emph{viz.} the members of the European Union in 2015 (see Table 1), considering two attributes: life expectancy at birth and GDP per capita, being measured in constant prices. The  data is listed in Tables 2 and 3, Appendix B. These attributes are incommensurable and necessitate a bivariate approach. As no common unit of measurement exists, any nonnegative vector $\bmp$ of relative attribute weights is considered.
Figure 1 exhibits $R_{DW}$-based PREs in the years 2000 and 2015, for various values of  $1/\alpha$. Figure 2  demonstrates their change over time for $1/\alpha=2$ and $1/\alpha=14/3$.

\begin{table}[h!]
\label{tab1}
\begin{center}
  \begin{tabular}{||l|l|r||l|l|r||}
  \hline
  country & abbr. & EU since & country & abbr. & EU since  \\
  \hline
Austria& AU & 1995 &
Belgium & BE & 1957\\
Bulgaria & BG & 2007 &
Croatia & CR & 2013\\
Cyprus & CY & 2004 &
Czech Republic & CZ & 2004\\
Denmark & DK & 1973 &
Estonia & ET & 2004\\
Finland & FI  & 1995 &
France & FR & 1957\\
Germany & GE  & 1957 &
Greece & GR  & 1981\\
 Hungary & HU & 2004 &
Ireland & IR & 1973\\
Italy & IT & 1957 &
Latvia & LT & 2004\\
Lithuania & LT & 2004 &
Luxembourg & LU  & 1957 \\
Malta & MA & 2004 &
Netherlands & NL & 1957 \\
Poland & PL & 2004 &
Portugal & PT & 1986\\
Romania & RO & 2007 &
Slovakia & SK & 2004\\
Slovenia & SN & 2004 &
Spain & SP & 1986\\
Sweden & SW & 1995 &
United Kingdom & UK & 1973\\
  \hline
  \end{tabular}
 \caption{European countries and year of accession to EU or its predecessors.}
\end{center}
\end{table}

\begin{figure}
\label{fig1}
	\includegraphics[width=0.48\textwidth]{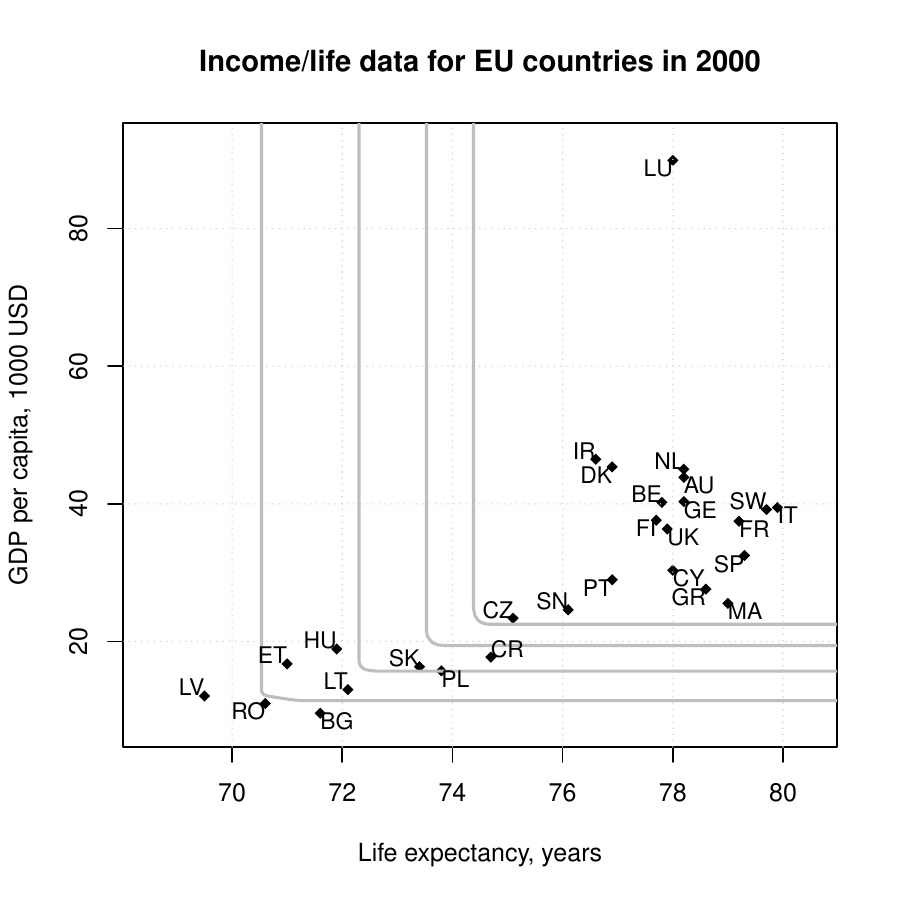}  \includegraphics[width=0.48\textwidth]{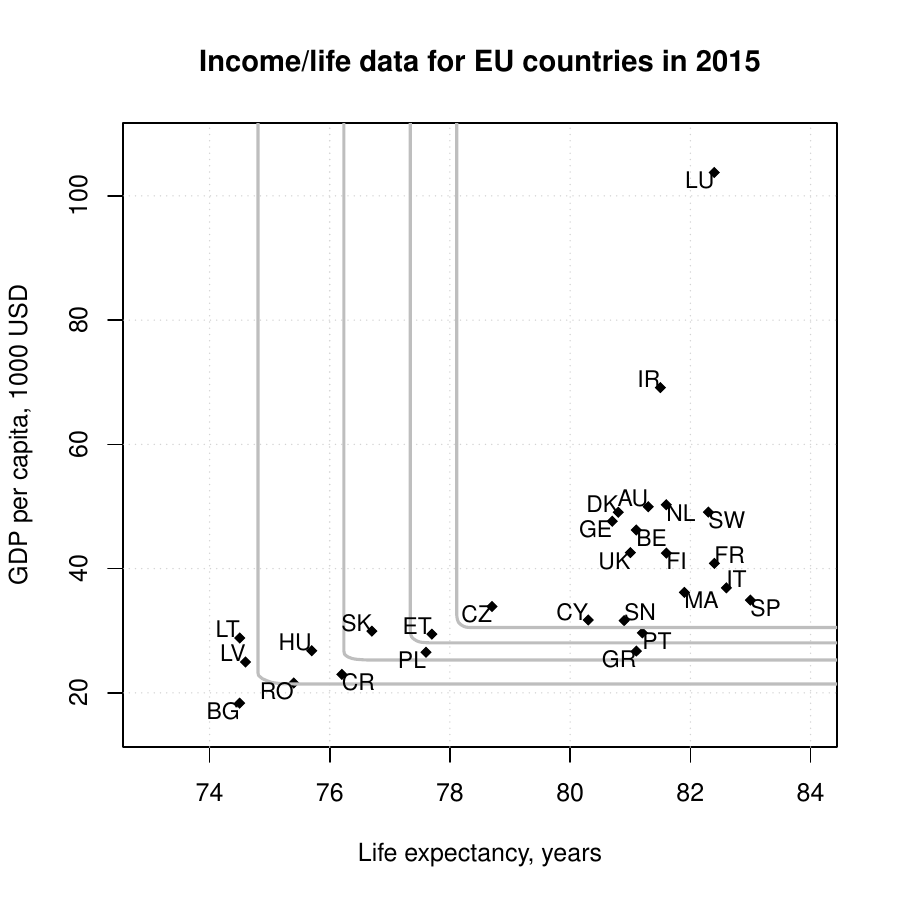}
      \caption{$R_{DW}$-based representative endowments of European countries (later EU-28) regarding life expectancy (years) and per capita GDP (1000 USD) in 2000 (left panel) and 2015 (right panel). The parameter is $1/\alpha= 2, 14/5, 14/3, 7$, lowest at the upper right curve and increasing to the lower left one.}
\end{figure}

\begin{figure}
\label{fig2}
\centering
	\includegraphics[width=0.6\textwidth]{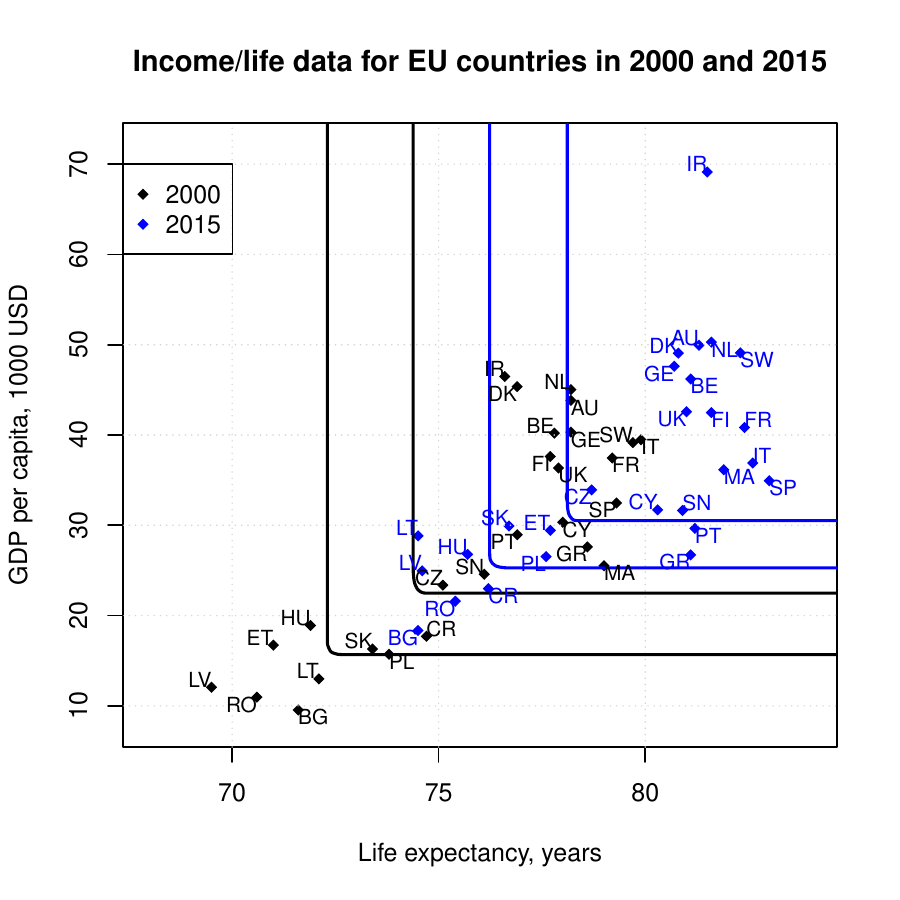}
      \caption{Comparison of $R_{DW}$-based representative endowments of European countries (EU-28) regarding life expectancy (years) and per capita GDP (1000 USD) in 2000 (black) and in 2015 (blue). The parameter is $1/\alpha= 2$ (upper right curves) resp. $14/3$ (lower left curves).}
\end{figure}

The two figures show that for these data all CREs are close to upper quadrants. The curved parts of them are rather small, which means that the effect of different weight vectors $\bmp$ is slight. Each PRE divides the countries into two groups, one above and one below the line. E.g. for $1/\alpha=2$ (upper right lines), which corresponds to the usual univariate Gini index, we see that in 2000 the countries below the PRE line consist of the East European states (besides Czechia and Slovenia), while all members of the European Union at that time are clearly above
the line. In 2015, after the Eastern countries entered the European Union, this divide remains (only Portugal and Greece move to the lower part), but the relative positions of Eastern countries changes, which is further described by PREs with lower degrees of $1/\alpha$.
Figure 2 emphasizes the movement of PREs over time: For each $\alpha$, the PRE in 2015 lies above that in 2000, which means that the distribution in 2015 has higher welfare regarding uniform Gini dominance than that in 2000.

While different classes of distortion functions provide different representative endowments, results appear to be  similar in a qualitative sense. Figure 3 in Appendix B shows the respective PREs based on the class of distortion functions (\ref{ralpha}).

\section{Scale invariant measurement of inequality}\label{sec:scale}
The multi-dimensioned representative endowment is a set-valued measure of welfare, reflecting both the inequality and the general level of the distribution. It is equivariant to translation and multivariate scale. However, to measure `pure' inequality usually scale invariant measures are preferred.
In this section we shortly discuss a scale invariant measure of inequality that is closely related to the above representative endowments.

Given a random vector $\bmX=(X_1,X_2,\dots,X_d)'$ in $\mathbb{R}^d$ having strictly positive expectation, $E[\bmX]=(\mu_1,\mu_2,\dots,\mu_d)' > 0$, denote its normalized version by
\[ \bmX^* = \left(\frac{X_1}{\mu_1}, \frac{X_2}{\mu_2}. \dots, \frac{X_d}{\mu_d}\right)'\,,
\]
and consider the convex representative endowment $C^+(\bmX^*, \tilde v)$ of the normalized $\bmX$ under weight-generating function $v$. This notion is scale invariant by construction.
A scale invariant ordering may be simply introduced as follows.

\begin{definition}
Given two random vectors $\bmX$ and $\bmY$ that have positive expectations,
define $\bmX\preceq^*_{vP} \bmY$ if
$\bmX^*\preceq_{vP} \bmY^*$. In words, $\bmY$ is \emph{less unequal than} $\bmX$ regarding distortion function $v\in V_{conc}$ and information $P$ on attribute weights.
\end{definition}
If $0<E[\bmX]= E[\bmY]=(\mu_1,\mu_2, \dots, \mu_d)'$ it follows from the definition that
\begin{equation}
\bmX\preceq^*_{vP} \bmY \;\; \text{if and only if} \;\; \bmX\preceq_{vP/\mu} \bmY\,,
\end{equation}
where $P/\mu = \{ \bmq=\gamma^{-1} (p_1/\mu_1,\ p_2/\mu_2, \dots, p_d/\mu_d)' : (p_1, p_2, \dots, p_d)' \in P, \gamma=\sum_{i=1}^d p_i/\mu_i\}$.
Since $P_+/\mu= P_+$, obtain:
\begin{prop}\label{pureinequality}
Let $E[\bmX]= E[\bmY]>0$ and $P=P_+$. Then
\[\bmX\preceq^*_{vP} \bmY \;\; \text{if and only if} \;\;\; \bmX\preceq_{vP} \bmY\,.
\]

\end{prop}
Among multi-dimensioned distributions of equal mean, the uniform Gini dominance $\preceq_{vP_+}$ is a scale invariant ordering of inequality.

\section{Ordering properties}\label{sec:Ordering}
The $P$-uniform Gini dominance, $\bmX\preceq_{vP} \bmY$, means that $\bmX$
has lower welfare than $\bmY$, given a function $v\in V_{conc}$ reflecting inequality posture and a set $P$ providing information on attribute weights. This Section presents aspects of $\preceq_{vP}$ as a reverse ordering of variability and its relations to known stochastic orderings of distributions.
Proposition \ref{VarOrder} states two properties of $\preceq_{vP}$ that are typical for a variability ordering:

\begin{prop}\label{VarOrder}
For any $v\in V_{conc}$ and $\emptyset \not= P\subseteq P_+$ it holds
\begin{itemize}
  \item [(i)] $\bmX \preceq_{vP}E[\bmX]$,
  \item [(ii)] $\beta(\bmX-E[\bmX]) \preceq_{vP} \bmX-E[\bmX] $ if $\beta\ge 1$.
\end{itemize}
\end{prop}
For proof see Appendix A. The Proposition says that the distribution of $\bmX$ has $vP$-lower welfare than the one-point distribution at $E[\bmX]$ and that,  for any $\beta\ge 1$, the centered distribution of $\bmX$ has $vP$-higher welfare than the same distribution `blown-up' by some factor $\beta>1$.
The next result is about the ordering of expectations.
\begin{prop}\label{ExpOrder}
$\bmX\preceq_{id P_+} \bmY \; \Rightarrow \; E[\bmX]\le E[\bmY],$ where $id$ is the identity function on $[0,1]$.
\end{prop}
For proof, note that when $v=id$, (\ref{uniformGini}) becomes $\int_0^1Q_{\bmp^\prime \bmX}(t) dt \le  \int_0^1Q_{\bmp^\prime \bmY}(t) dt$ for all
$\bmp \ge 0$, which implies $E[\bmX]\le E[\bmY]$.

Now, let us compare uniform Gini dominance relations with known stochastic orders.
A random vector $\bmX$ in $\mathbb{R}^d$ is dominated by another random vector $\bmY$ in
\textit{increasing concave (convex) order} $\ge_{iconc}$ (resp.\ $\ge_{iconv}$), if
\begin{equation}\label{intorder}
E[g(\bmX)] \le E[g(\bmY)]
\end{equation}
holds for all functions $g:\IR^d \to \IR$ that are componentwise increasing and concave (resp.\ convex) and for which both expectations exist.
$\bmY$ is \textit{stochastically larger} than $\bmX$, $\bmX\preceq_{st}\bmY$, if (\ref{intorder}) is satisfied for all increasing $g$.
Obviously, $\preceq_{st}$ is stronger than $\preceq_{iconc}$. The latter proves to be stronger than uniform Gini dominance, for any $v\in V_{conc}$ and $P\subseteq P_+$:

\begin{prop}\label{iconcOrder}
It holds that $\bmX\preceq_{iconc} \bmY$ implies
\begin{center}
$\bmX\preceq_{vP} \bmY$ for all $v\in V_{conc}$ and $\emptyset \not= P\subseteq P_+$.
\end{center}
\end{prop}

For proof, see Appendix A.

As we have seen, the Gini dominance orders reflect the spread of the attributes' distributions as well as their general levels. Further, they indicate differences in correlation. For example, consider bivariate random vectors, $\bmX=(X_1,X_2)'$ and $\bmY=(Y_1,Y_2)'$, each being concentrated at two points with equal probabilities,
\[ P[\bmX=(0,0)']=P[\bmX=(1,1)']= 0.5\,,\ P[\bmY=(1,0)']=P[\bmY=(0,1)']= 0.5\,.\]
With $\bmX$ one unit has high, the other unit low value in both attributes, while with $\bmY$ each unit has one high and one low value. These distributions have identical marginals, and $\bmX$ is more correlated than $\bmY$. Then for any $\bmp=(p_1,p_2)'\in P \subseteq P_+$,
\[\int_0^1 Q_{\bmp'\bmX}(t) dv(t)= v(0.5) \bmp'(0,0)' + (v(1) -v(0.5)) \bmp'(1,1)' =  (1 -v(0.5)) (p_1 + p_2) = 1 -v(0.5)\,.\]
In case $0<p_1\le p_2$ it holds
\[\int_0^1 Q_{\bmp'\bmY}(t) dv(t)= v(0.5)p_1 + (1 -v(0.5))p_2\,,\]
and we obtain
$\bmX\preceq_{v \{p\}} \bmY$ if and only if $1 -v(0.5) \le  v(0.5) p_1 +  (1 -v(0.5))p_2$ if and only if $v(0.5) \ge 0.5,$ which is true since $v\in V_{conc}$.
In case $0<p_2 \le p_1$, we similarly get $\bmX\preceq_{v \{p\}} \bmY$ if and only if $v(0.5) \ge 0.5.$ Consequently, $\bmX\preceq_{v \{p\}} \bmY$ holds for every $\bmp\in P_+$, hence $\bmX\preceq_{vP_+} \bmY$: $\bmY$ has higher $v$-welfare in uniform Gini dominance than $\bmX$.
Here, under equal marginals the higher correlation of $\bmX$ results in less welfare .

General orderings of dependence are the submodular order and the concordance order; see e.g.\ \cite{MuellerS02}. A function $f:\IR^d \to \IR$ is \textit{submodular} if
\begin{equation}\label{submodular}
f(\bmx \wedge \bmy) + f(\bmx \vee \bmy) \le f(\bmx) + f(\bmy)\quad \text{for all} \; \bmx, \bmy \,.
\end{equation}
$\bmY$ is \textit{larger than $\bmX$ in submodular order}, $\bmX\preceq_{submod} \bmY$,  if (\ref{intorder}) holds for all submodular functions (\ref{submodular}) as far as the expectations exist. Like the increasing concave order, the submodular order is stronger than the ordering $\preceq_{vP}$ (for proof see Appendix A):
\begin{prop}\label{submodOrder}
It holds that $\bmX\preceq_{submod} \bmY$ implies
\begin{center}
$\bmX\preceq_{vP} \bmY$ for all $v\in V_{conc}$ and $\emptyset \not= P\subseteq P_+$.
\end{center}
\end{prop}

The \textit{concordance order}, $\bmX\preceq_{c} \bmY$, says that for each $i,j \in \{1,2,\dots, d\}$ the bivariate distributions $F_{\bmX_{ij}}$ and $F_{\bmY_{ij}}$ are pointwise ordered, $F_{\bmX_{ij}}(\bmz) \le F_{\bmY_{ij}}(\bmz)$ for all $\bmz\in \IR^2$. This implies equal marginals and increasing covariances, $\Cov(X_i,X_j) \le \Cov(Y_i,Y_j)$. The reverse concordance order is weaker than the submodular order, $\bmY\preceq_{c} \bmX$ follows from $\bmX\preceq_{submod} \bmY$.

In view of the above one may consider Gini dominance relations that are uniform not only in attribute weights, \emph{viz.}\ the vector $\bmp$,   but also in their attitude to inequality aversion, which is described by some family of functions $v$.

\begin{definition}
Given some convex-ordered family $R$, define $\bmX\preceq_{RP} \bmY$, which means ${P}$-uniform Gini dominance for all $v$ in $R\subseteq V_{conc}$, in words, $\bmX$ has \emph{higher $RP$-welfare} than $\bmY$.
\end{definition}
This kind of Gini dominance is doubly uniform in attribute weights as well as in distortion functions of different inequality aversion collected in $R$.
From Propositions \ref{iconcOrder} and \ref{submodOrder} follows that, for any $R\subseteq V_{conc}$, $\bmX$ has lower $RP$-welfare than $\bmY$ if either $\bmX$ is less than $\bmY$ in increasing concave stochastic order or in submodular order.

Consider the special case $R=R_{zon}$. The positive weighted-mean order $\preceq_{R_{zon} P_+}$, based on the zonoid regions, is also known as
\emph{weak price submajorization}; see \cite[Sec. 9.4]{Mosler02a}. For a general convex-ordered family $R$ and given $P$ the relation $\preceq_{RP}$ may be mentioned as \emph{$P$-positive weighted-mean order} as it is the variant of a weighted-mean order \citep{DyckerhoffM11} restricted to non-negative directions in $P$.
It can be shown that the ordering $\preceq_{R_{zon} P}$ implies any other $P$-positive weighted-mean order:
\begin{prop}\label{orderZonoid}
Let $R\subseteq V_{conc}$ be convex ordered and $\emptyset \not= P\subseteq P_+$.
Then
\[\bmX \preceq_{R_{zon}P} \bmY \;\; \Rightarrow \;\; \bmX \preceq_{RP}\bmY\,.
\]
\end{prop}
The proof is similar to that of Proposition 11 in \cite{DyckerhoffM10a}, who also provide examples of weighted-mean orderings $\preceq_{RP}$ that are different from $\preceq_{R_{zon}P}$.
The following theorem collects the above results regarding orders that are stronger than $\preceq_{RP}$, that is, imply $\preceq_{RP}$.

\begin{theorem}\label{convorder}
Let $R \subseteq V_{conc}$ be convex ordered and $\emptyset \not= P\subseteq P_+$. Sufficient for $\bmX \preceq_{RP} \bmY$  is each of the following restrictions:
\begin{enumerate}
\item[(i)] $\bmX\preceq_{R_{zon}P} \bmY$\,,
\item[(ii)] $\bmX\preceq_{iconc} \bmY$\,,
\item[(iii)] $\bmX\preceq_{conc} \bmY$\,,
\item[(iv)] $\bmX\preceq_{st} \bmY$\,,
\item[(v)] $\bmX\preceq_{submod} \bmY$\,,
\end{enumerate}
and the reverse implications are {generally wrong}.
\end{theorem}

Specifically consider two Gaussian vectors  $\bmX$ and $\bmY$, $\bmX \sim N(\mu_\bmX, \Sigma_\bmX)$, $\bmY \sim N(\mu_\bmY, \Sigma_\bmY)$.
Then the mentioned orders are characterized by first and second moments as follows, which yields parametric conditions that are sufficient for $\bmX \preceq_{RP} \bmY$.

\begin{prop}\label{orderedNormals}
\begin{itemize}
  \item[(i)] $\bmX\preceq_{R_{zon}P} \bmY \quad \Leftrightarrow  \quad \mu_\bmX \le \mu_\bmY \;\; \text{and} \;\; \bmp'(\Sigma_\bmX - \Sigma_\bmY)\bmp\ge 0 \;\; \text{for all} \;\; \bmp\in P,$
  \item[(ii)] $\bmX\preceq_{iconc} \bmY \quad  \Leftrightarrow \quad \mu_\bmX \le \mu_\bmY \;\; \text{and}\;\;\Sigma_\bmX - \Sigma_\bmY \;\; \text{nonnegative definite},$
  \item[(iii)] $\bmX\preceq_{conc} \bmY \quad  \Leftrightarrow \quad \mu_\bmX = \mu_\bmY \;\; \text{and} \;\; \Sigma_\bmX - \Sigma_\bmY \;\; \text{nonnegative definite},$
  \item[(iv)] $\bmX\preceq_{st} \bmY \quad  \Leftrightarrow \quad \mu_\bmX \le \mu_\bmY \;\; \text{and}\;\; \Sigma_\bmX = \Sigma_\bmY,$
\item[(v)] $\bmX\preceq_{submod} \bmY \quad  \Leftrightarrow \quad \bmY\preceq_{c} \bmX \quad \Leftrightarrow \quad \mu_\bmX = \mu_\bmY\,, \;\; \Var(X_i)=\Var(Y_i) \;\; \text{and}\;\; \Cov(X_i,X_j) \ge \Cov(Y_i,Y_j) \;\; \text{for all}\; i,j.$
\end{itemize}
\end{prop}
For proof see Appendix A.

\section{Statistical and computational issues}\label{sec:Statistical}
For the comparison of welfare in multiple attributes
set-valued representative endowments have been introduced as well as $P$-uniform Gini dominance orderings.
We conclude the paper by some remarks on computational and statistical issues that arise when these notions are used with data.
Consider data in $\IR^d$, $d\ge 2$, that is, an empirically distributed $\bmX$. To numerically determine the convex representative endowment  $C^+(\bmX,\tilde v)$ for a given concave function $v$ that generates the inequality weights, the exact algorithm of \cite{BazovkinM12} and the R-package \cite{Bazovkin13} may be used.
This procedure calculates the set $C(\bmX,\tilde v)$, being a so called weighted-mean (WM) region. Its lower boundary coincides with the lower boundary
of $C^+(\bmX,\tilde v)$, which corresponds to attribute weight vectors $\bmp \in P_+$. It constructs the WM region \textit{by its facets}, that is, step-by-step building the surface of the convex polytope in $\IR^d$.
The algorithm is easily restricted to the part of the lower boundary that corresponds to a specified set $P$ of attribute weights $\bmp$.

In a sampling context one may ask whether a sample of $d$-variate representative endowments converges to the representative endowment of the underlying probability distribution.
Consider an i.i.d.\ sample $\bmX_1, \bmX_2, \dots, \bmX_n,\dots $ from $\bmX$ in $\IR^d$ and the sequence of representative endowments $C^+(\bmX_n, \tilde w_n)$.
\cite{DyckerhoffM11} have shown a strong law of large numbers: If $\tilde w_n(t)\in V_{conv}$ converges pointwise to some $\tilde w(t)$ and $\sup_n \tilde w'_n(1)<\infty$, then $C(\bmX_n, \tilde w_n)$ converges to $C(\bmX, \tilde w)$ in Hausdorff metric with probability one, which is tantamount saying that
\[ Prob\, [ \lim_n h_{C(\bmX_n, \tilde w_n)}(\bmp) = h_{C(\bmX, \tilde w)}(\bmp)\quad \text{for all} \;\; \bmp\in \mathbb{R}^d]=1\,.
\]
This convergence result implies the Hausdorff convergence of $C^+(\bmX_n, \tilde w_n)$ to $C^+(\bmX, \tilde w)$. We conclude that the empirical PRE is a strongly consistent estimator of its theoretical counterpart $C^+(\bmX, \tilde w)$.

\section*{Data} All data analysed in this article are from open sources (OECD and UN Statistics).

\section*{Acknowledgements} I'm greatly indebted to Pavlo Mozharovskyi for preparing the figures and to Friedrich Schmid for his remarks on a previous version of the paper. I also thank two anonymous referees for their most diligent reading and many valuable suggestions that added significantly to the paper.




\section*{Appendix A}
\begin{lemma}\label{lemmadualsupport}
Let $v\in V_{conc}$ and $\tilde v$ denote its \textit{dual distortion function},
$\tilde v(t) = 1- v(1-t)$. Then
for any $\bmp, \bmb \in \IR^d$ it holds
\[ S_v(\bmp'(\bmX+\bmb))= \int_0^1Q_{\bmp^\prime (\bmX+\bmb)}(t)\,dv(t) = - h_{C(\bmX -\bmb,\tilde v)}(-\bmp)\,.
\]
\end{lemma}
\textbf{Proof:}
\begin{eqnarray*}
S_v(\bmp'(\bmX+\bmb))&=&  \int_0^1Q_{\bmp^\prime (\bmX+\bmb)}(t)\,dv(t) \nonumber \\
  &=& \int_0^1[Q_{\bmp^\prime \bmX}(t)+\bmp'\bmb]\,d(1-\tilde v(1-t)) \nonumber\\
  &=& \int_0^1 [Q_{\bmp^\prime \bmX}(1-s)+\bmp'\bmb]\,d\tilde v(s)\nonumber\\
  &=& - \int_0^1[Q_{-\bmp^\prime \bmX}(s)+\bmp'\bmb]\,d\tilde v(s) \nonumber\\
  &=& - \int_0^1 Q_{-\bmp^\prime (\bmX- \bmb)}(s)\,d\tilde v(s) \nonumber\\
    &=& - h_{C(\bmX-\bmb,\tilde v)}(-\bmp)\,.
\end{eqnarray*}
\qed

\textbf{Proof of Proposition \ref{ExtremalP}:}
$\Leftarrow$: Let $\bmp\in P$. It is a convex combination, $\bmp = \sum_{j=1}^m \lambda_j \bmq_j$, of points $q_j\in Ext(P)$.
From $\bmX\preceq_{v Ext(P)} \bmY$ follows
\begin{eqnarray}
\int_0^1 Q_{\bmq_j'\bmX}(t) dv(t) &\le & \int_0^1 Q_{\bmq_j'\bmY}(t) dv(t)\,\;\; \text{for $j=1,\dots,m$, hence} \nonumber\\
\int_0^1 \lambda_j Q_{\bmq_j'\bmX}(t) dv(t) &\le & \int_0^1 \lambda_j Q_{\bmq_j'\bmY}(t) dv(t)\, \;\;\text{for $j=1,\dots,m$, \,and} \nonumber\\
\sum_{j=1}^m \int_0^1 \lambda_j Q_{\bmq_j'\bmX}(t) dv(t) &\le & \sum_{j=1}^m \int_0^1 \lambda_j Q_{\bmq_j'\bmY}(t) dv(t)
\,.  \nonumber
\end{eqnarray}
As $\{\bmp'\bmX : \bmp\in Ext(P)\}$ is comonotonic, the quantile values $Q_{\bmq_j'\bmX}(t)$ are ordered in the same way for all $j=\{1,\dots, m\}$ and thus receive the same inequality weights. Therefore
\begin{eqnarray*} \sum_{j=1}^m \int_0^1 \lambda_j Q_{\bmq_j'\bmX}(t) dv(t)& = & \int_0^1 \sum_{j=1}^m \lambda_j Q_{\bmq_j'\bmX}(t) dv(t) \\
&=& \int_0^1 Q_{\bmp'\bmX}(t) dv(t)\,,
\end{eqnarray*}
and the same for $\bmp'\bmY$. We conclude
\[ \int_0^1 Q_{\bmp'\bmX}(t) dv(t) \le  \int_0^1 Q_{\bmp'\bmY}(t) dv(t)\,. \label{all_p}\]
It holds (\ref{all_p}) for all $\bmp\in P$, that is $\bmX\preceq_{v P} \bmY$ \,. \\
$\Rightarrow$: Obvious since $Ext(P) \subseteq P$. \qed

\textbf{Proof of Proposition \ref{affinetransform}:}
Let $A\in \IR_+^{m\times d}$, $\bmb \in \IR^m$, and assume $\bmp\in \IR^m_+$.
Then we have $A'\bmp \ge 0$ and, by (\ref{support+}) and (\ref{dualsupport}),
\begin{eqnarray}
  h_{C^+(A\bmX+\bmb,\tilde v)}(-\bmp) &=& h_{C(A\bmX+\bmb,\tilde v)}(-\bmp) \nonumber\\
  &= &  - \int_0^1 Q_{\bmp' (A\bmX+\bmb)}(t)\,d  v(t)\,,\quad  \text{further}  \nonumber\\
  & =& - \int_0^1 Q_{(A'\bmp)'\bmX+\bmp'\bmb}(t)\,d  v(t) \nonumber \\
  &=&  - \int_0^1 Q_{(A'\bmp)'\bmX}(t)\,d  v(t) - \bmp'\bmb \nonumber\\
  &=&   h_{C^+(\bmX,\tilde v)}(-A'\bmp) - \bmp'\bmb \label{proofProp2}\\
  &=&   h_{AC^+(\bmX,\tilde v)+\bmb}(-\bmp) \label{proofProp2'}\,.
\end{eqnarray}
Here (\ref{proofProp2}) follows from Lemma \ref{lemmadualsupport} applied to $A'\bmp$ in place of $\bmp$ and $\bmX + \bmb$ in place of $\bmX$, while  (\ref{proofProp2'}) holds since
$h_K(A'\bmp)+\bmp'\bmb =h_{AK+\bmb}(\bmp)$ for any closed convex set $K$.
If $\bmp\not\in P_+$, obtain $h_{C^+(A\bmX+\bmb,\tilde v)}(-\bmp) =\infty$ as well as $h_{AC^+(\bmX,\tilde v)}(-\bmp)=
h_{AC^+(\bmX,\tilde v)+\bmb}(-\bmp)=\infty$. We conclude $h_{C^+(A\bmX+\bmb,\tilde v)}(\bmp)=  h_{AC^+(\bmX,\tilde v)+\bmb}(\bmp)$ for all $\bmp\in \mathbb{R}^d$, which yields the claim  (\ref{affineequivariant}).\qed

 \textbf{Proof of Proposition \ref{VarOrder}:}
(i): Seen as probability distribution functions, the concave function $v$ is dominated by the identity function $w(t)=t$ in first degree stochastic dominance. Hence
$\int_0^1 \phi(t) dt \ge \int_0^1 \phi(t) dv(t)$ for any increasing function $\phi$, in particular, for the quantile function
$\phi(t) =Q_{\bmp'\bmX} (t)$. It follows, for all $\bmp\in \mathbb{R}^d$,
\begin{equation}\label{ExpInclusion}
\int_0^1 Q_{\bmp'E[\bmX]} (t) d v(t)=\bmp'E[\bmX]
=\int_0^1 Q_{\bmp'\bmX} (t) dt \ge \int_0^1 Q_{\bmp'\bmX} (t) d v(t)\,.
\end{equation}
We conclude $\bmX \preceq_{vP} E[\bmX] $ for any $\emptyset \not= P\subseteq P_+$.

(ii): Again, let $v\in V_{conc}$ and  $\bmp \in P \subseteq P_+$.
As in (i) it follows that
\[  \int_0^1 Q_{\bmp'(\bmX-E[\bmX])} (t) dv(t) \le \int_0^1 Q_{\bmp'(\bmX-E[\bmX])} (t) dt = 0\,.\]
If $\beta \ge 1$ we get
\[ \int_0^1 Q_{\bmp'\beta(\bmX-E[\bmX])} (t) dv(t) = \beta \int_0^1 Q_{\bmp'(\bmX-E[\bmX])} (t) dv(t) \le \int_0^1 Q_{\bmp'(\bmX-E[\bmX])} (t) dv(t)\,.
\]
\qed

\textbf{Proof of Proposition \ref{iconcOrder}:}
 Let $\bmX\preceq_{iconc} \bmY$ and $g:\mathbb{R} \to \mathbb{R}$ be increasing and concave. Then for any $\bmp\ge 0$ the function $\bmx\mapsto g(\bmp'\bmx),$ $\bmx \in \mathbb{R}^d$, is also increasing and concave, and
$E[g(\bmp'\bmX)] \le E[g(\bmp'\bmY)]$ holds by (\ref{intorder}). It follows that $\bmp'\bmX \preceq_{iconc} \bmp'\bmY$ in univariate increasing concave order, which is the same as  $-\bmp'\bmY \preceq_{iconv} - \bmp'\bmX$ in univariate increasing convex order. By Theorem 4.A.4 in \cite{ShakedS07} this holds if and only if
\begin{eqnarray*}
\int_0^1 Q_{-\bmp'Y} (t)dw(t) & \le & \int_0^1 Q_{-\bmp'X} (t)dw(t)\, \;\; \text{for all} \;\; w\in V_\conv\,, \quad \text{which is equivalent to}\\
\int_0^1 Q_{\bmp'\bmX} (t)dv(t) & \le & \int_0^1 Q_{\bmp'Y} (t)dv(t)\, \;\; \text{for all} \;\; v\in V_{conc}\,.
\end{eqnarray*}
As this is true for every $\bmp\ge 0$, we obtain $\bmX \preceq_{vP} \bmY$ for $v\in V_{conc}$ and  $P\subseteq P_+$.\qed

\textbf{Proof of Proposition \ref{submodOrder}:}
A function $f$ is \textit{supermodular} if it satisfies (\ref{submodular}) with reverse inequality sign.
The \textit{supermodular order} $\preceq_{supermod}$ is defined by (\ref{intorder}) for all supermodular $f$.
Note that if $f$ is submodular the function $g:\bmx\mapsto -f(-\bmx)$ is supermodular, and \textit{viceversa}.
Now, assume $\bmX\preceq_{submod}\bmY$, that is, $E[f(\bmX)]\le E[f(\bmY)]$ for all submodular $f$. Then
$-E[f(-\bmX)] \le -E[f(-\bmY)]$ for all supermodular $f$, that is, $-\bmY \preceq_{supermod} -\bmX$ is satisfied.
It follows from Theorem 3.9.5(d) in \cite{MuellerS02} that
 $-\bmp'\bmY \preceq_{conv} -\bmp'\bmX$ holds for all $\bmp\ge 0$, hence $-\bmp'\bmY \preceq_{iconv} -\bmp'\bmX$. As in the proof of Proposition \ref{iconcOrder} conclude
\begin{eqnarray*}
\int_0^1 Q_{ - \bmp'\bmY} (t)dw(t) & \le & \int_0^1 Q_{- \bmp'X} (t)dw(t)\,, \quad v\in V_{conv}\,,\\
 \int_0^1 Q_{\bmp'X} (t)dv(t) & \le & \int_0^1 Q_{\bmp'Y} (t)dv(t)\,, \quad v\in V_{conc}\,,
\end{eqnarray*}
that is, $\bmX \preceq_{vP} \bmY$ for any $v\in V_{conc}$.
\qed

\textbf{Proof of Proposition \ref{orderedNormals}:}
(i): Can be concluded from \cite[Proposition 8.20]{Mosler02a}.\\
(ii) and (iii): See e.g. \cite[Corollary 3.2]{Mosler84}.\\
(iv): See e.g. \cite[Theorem 3.1]{Mosler84}.\\
(v): See Example 3.8.6. and Theorem 3.13.5 in \cite{MuellerS02}. \qed

\section*{Appendix B}

This appendix contains the data of the numerical example (Tables 2 and 3)  and the pertaining representative endowments based on the class (\ref{ralpha})
of distortion functions (Figure 3).

\begin{table}[h!]\label{tab2}
\begin{center}
  \begin{tabular}{||l|l|r|r||l|l|r|r||}
  \hline
  2000 &&&& 2000 &&& \\
  \hline
  country & & life exp. & GDP & country & & life exp. & GDP  \\
  \hline
Austria & AU & 78.2 & 43826 &
Belgium & BE &  77.8 & 40204\\
Bulgaria &  BG & 71.6 & 9537 &
Croatia & CR & 74.7 & 17707\\
Cyprus & CY & 78.0 & 30338 &
Czech Republic & CZ & 75.1 & 23370 \\
Denmark & DK &76.9 & 45363 &
Estonia & ET & 71.0 & 16719  \\
Finland & FI & 77.7 & 37615  &
France & FR & 79.2 & 37450 \\
Germany &  GE & 78.2 & 40320 &
Greece &  GR & 78.6 & 27608\\
 Hungary & HU & 71.9 & 18897  &
Ireland &  IR & 76.6 & 46480 \\
Italy &  IT & 79.9 & 39472  &
Latvia & LT & 69.5 & 12061\\
Lithuania & LT &  72.1 & 12985 &
Luxembourg & LU & 78.0 & 89924  \\
Malta & MA & 79.0 & 25525 &
Netherlands & NL & 78.2 & 45017  \\
Poland & PL & 73.8 & 15712  &
Portugal & PT & 76.9 & 28960 \\
Romania & RO & 70.6 & 10961  &
Slovakia & SK & 73.4 & 16303 \\
Slovenia & SN & 76.1 & 24583&
Spain &  SP & 79.3 & 32468\\
Sweden & SW & 79.7 & 39169  &
United Kingdom & UK & 77.9 & 36344 \\
  \hline
  \end{tabular}
 \caption{Life expectancy (years) and per capita GDP (1000 USD) of 28 European countries (later EU-28) in 2000. Source: OECD and UN.}
\end{center}
\end{table}

\begin{table}[h!]\label{tab3}
\begin{center}
  \begin{tabular}{||l|l|r|r||l|l|r|r||}
  \hline
  2015 &&&& 2015 &&& \\
  \hline
  country & & life exp. & GDP & country & & life exp. & GDP  \\
  \hline
Austria & AU &  81.3 & 49942  &
Belgium & BE &  81.1 & 46214\\
Bulgaria &  BG & 74.5 & 18343  &
Croatia & CR & 76.2 & 22981\\
Cyprus & CY & 80.3 & 31714 &
Czech Republic & CZ & 78.7 & 33909\\
Denmark & DK &80.8 & 49058  &
Estonia & ET & 77.7 & 29436 \\
Finland & FI & 81.6 & 42490  &
France & FR & 82.4 & 40830\\
Germany &  GE & 80.7 & 47610  &
Greece &  GR & 81.1 & 26721 \\
 Hungary & HU & 75.7 & 26777  &
Ireland &  IR & 81.5 & 69134\\
Italy &  IT & 82.6 & 36899  &
Latvia & LT & 74.6 & 24964\\
Lithuania & LT &  74.5 & 28834  &
Luxembourg & LU & 82.4 & 103760  \\
Malta & MA & 81.9 & 36157  &
Netherlands & NL & 81.6 & 50288  \\
Poland & PL & 77.6 & 26535   &
Portugal & PT & 81.2 & 29661 \\
Romania & RO & 75.4 & 21599  &
Slovakia & SK & 76.7 & 29928  \\
Slovenia & SN & 80.9 & 31632 &
Spain &  SP & 83.0 & 34929\\
Sweden & SW & 82.3 & 49103  &
United Kingdom & UK & 81.0 & 42572 \\
  \hline
  \end{tabular}
 \caption{Life expectancy (years) and per capita GDP (1000 USD) of 28 European countries (EU-28) in 2015. Source: OECD and UN.}
\end{center}
\end{table}

\begin{figure}
\label{fig3}
	\includegraphics[width=0.48\textwidth]{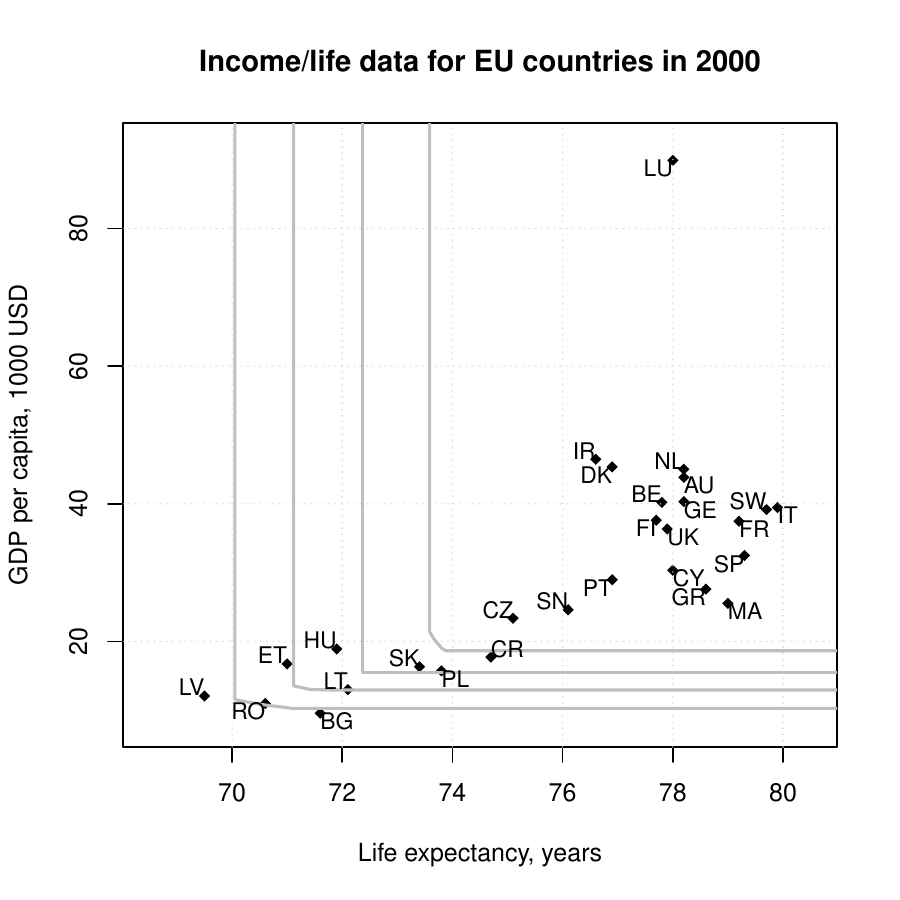}  \includegraphics[width=0.48\textwidth]{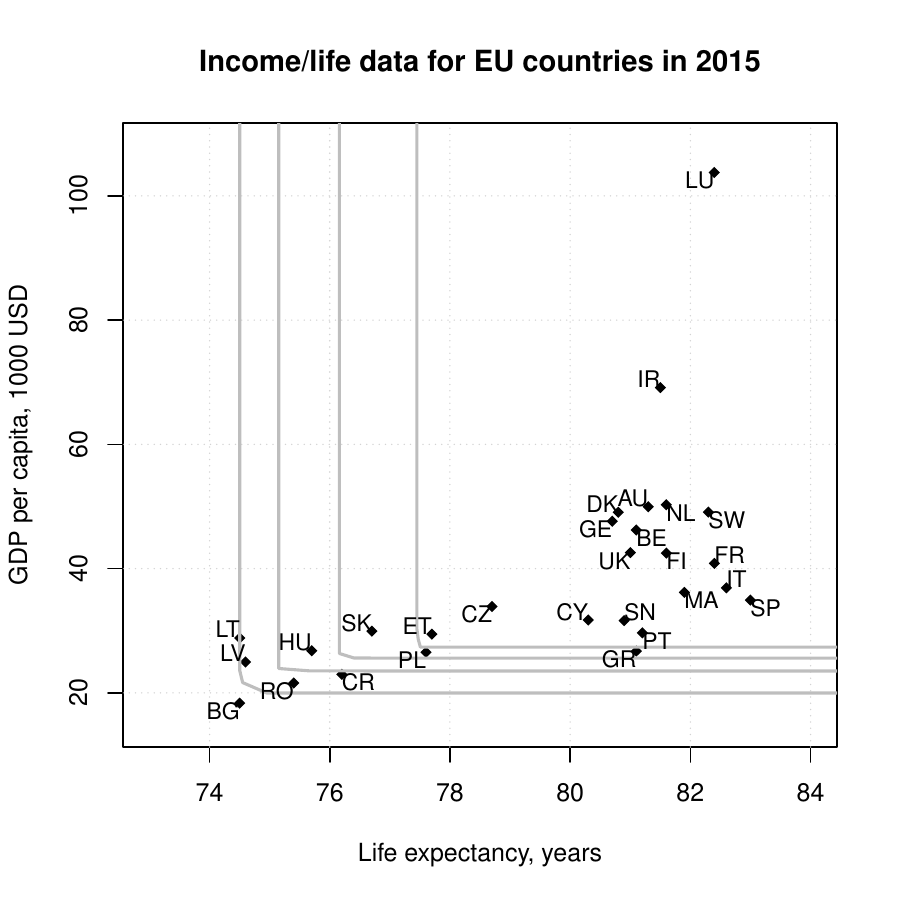}
      \caption{
      Representative endowments of European countries regarding life expectancy (years) and per capita GDP (1000 USD) in 2000 (left panel) and 2015 (right panel) based on the class (\ref{ralpha}) of distortion functions. The parameter is $1/\alpha= 2, 14/5, 14/3, 7$, lowest at the upper right curve and increasing to the lower left one.}
      \end{figure}

\end{document}